\documentclass[a4paper,fleqn]{cas-dc}
\usepackage{gensymb}
\usepackage{float}
\usepackage{amsthm}
\usepackage{siunitx}

\usepackage[numbers,sort&compress]{natbib}
\def\tsc#1{\csdef{#1}{\textsc{\lowercase{#1}}\xspace}}
\tsc{WGM}
\tsc{QE}
\tsc{EP}
\tsc{PMS}
\tsc{BEC}
\tsc{DE}

\begin{document}
\let\WriteBookmarks\relax
\def\floatpagepagefraction{1},
\def\textpagefraction{.001}
\shorttitle{Intermediate-energy proton irradiation}
\shortauthors{S Jepeal}

\title [mode = title]{Intermediate energy proton irradiation: rapid, high-fidelity materials testing for fusion and fission energy systems}

\author[1]{Jepeal S. J.}
\author[2]{Snead L.}
\author[1]{Hartwig. Z. S.}

\address[1]{Department of Nuclear Science and Engineering, Massachusetts Institute of Technology}
\address[2]{Materials Science and Chemical Engineering, Stony Brook University}

\begin{abstract}
Fusion and advanced fission power plants require advanced nuclear materials to function under new, extreme environments. Understanding the evolution of mechanical and functional properties during radiation damage is essential to the design and  commercial deployment of these systems. The shortcomings of existing methods could be addressed by a new technique - intermediate energy proton irradiation (IEPI) - using beams of 10 - 30 MeV protons to rapidly and uniformly damage bulk material specimens before direct testing of engineering properties. IEPI is shown to achieve high fidelity to fusion and fission environments in both primary damage production and transmutation, often superior to nuclear reactor or typical (low-range) ion irradiation. Modeling demonstrates that high dose rates (0.1--1~DPA/per day) can be achieved in bulk material specimens (100--\SI{300}{\micro\meter}) with low temperature gradients and induced radioactivity. The capabilities of IEPI are demonstrated through a 12 MeV proton irradiation and tensile test of \SI{250}{\micro\meter} thick tensile specimens of a nickel alloy (Alloy 718), reproducing neutron-induced data. These results demonstrate that IEPI enables high throughput assessment of materials under reactor-relevant conditions, positioning IEPI to accelerate the pace of engineering-scale radiation damage testing and allow for quicker and more effective design of nuclear energy systems.

\end{abstract}

\begin{graphicalabstract}
\includegraphics[width=.96\textwidth]{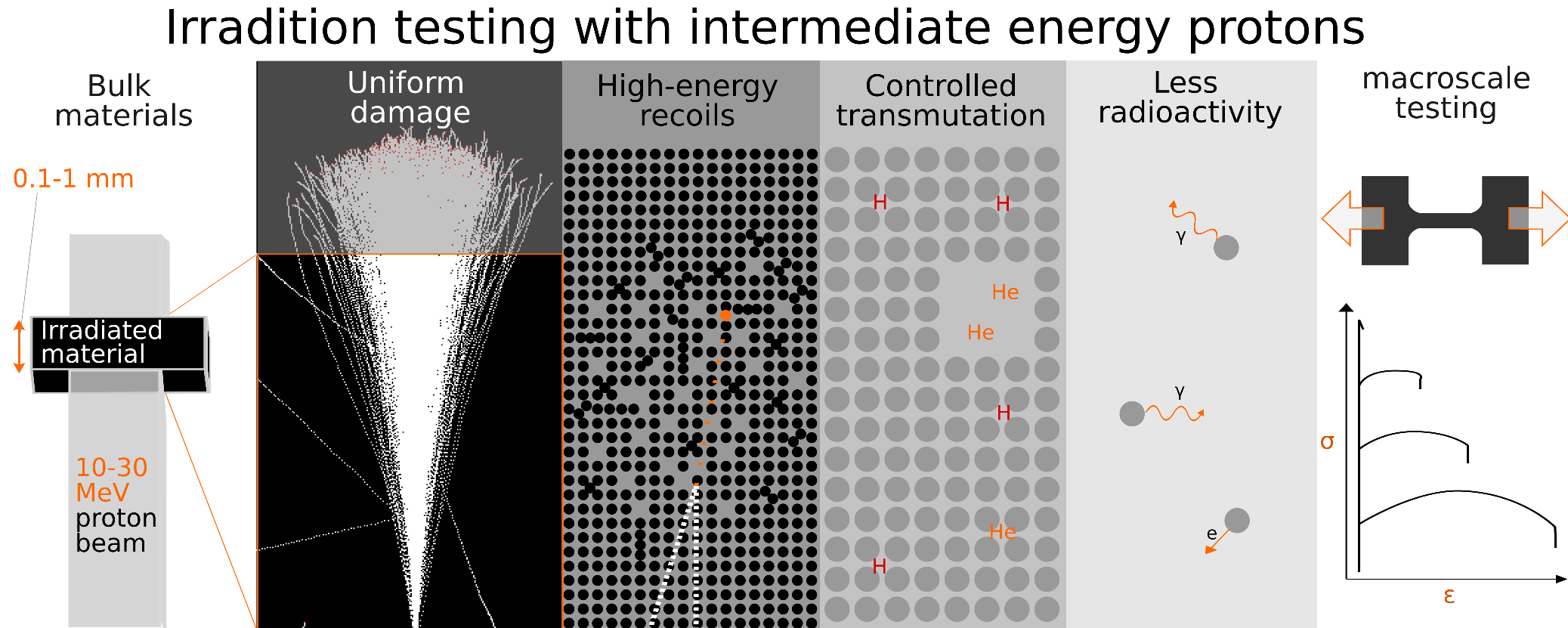}
\end{graphicalabstract}

\begin{highlights}
\item Irradiation of tungsten by 5-40 MeV protons generate recoils with energies of $10^{2}-10^{6}$~eV and a weighted mean of $\sim10^{5}$~ev, matching recoils from fission and fusion reactor neutron spectra.  
\item Irradiation of iron, tungsten, nickel, and copper by 5--20 MeV protons will directly generate helium transmutation, up to approximately $10^{2}$ app/dpa, and equivalent to irradiation a fusion power neutron spectrum.   
\item Intermediate-energy proton irradiation can achieve 0.1--1~dpa/day dose-rates in 100--\SI{300}{\micro\meter} samples without substantial temperature gradients. 
\item 12 MeV proton irradiation of \SI{250}{\micro\meter} thick alloy 718 produces bulk irradiation hardening in agreement with reactor irradiation as measured through tensile tests.

\end{highlights}


\maketitle

\section{Introduction: opportunities for intermediate-energy proton irradiation}
\label{Sec:Introduction}
Future fusion power plants and advanced nuclear fission reactors are being pursued as clean and affordable energy sources to mitigate global climate change. Fully realizing the potential of these technologies will require the design and qualification of materials that satisfy demanding operational requirements while surviving in extreme environments~\cite{Chant2010,Ehrlich2000,Linsmeier2017}. One of the most pressing challenges to these systems is the expected long term exposure of  structural and functional materials to extreme radiation fluxes. For example, materials surrounding the core of a fusion power plant are expected to receive total neutron fluxes of $10^{14}$-$10^{15}$ cm$^{-2}$ s$^{-1}$, equivalent to up to 15 displacements/atom (DPA) per year~\cite{Gilbert2012}. Such high levels of neutron exposure will cause undesirable property changes such as hardening, embrittlement, and decreases in thermal conductivity~\cite{Katoh2019}. For the effective design of future fission and fusion power plants, the relevant radiation damage effects on materials must be understood and mitigated. ~\cite{Maisonnier2018,Wu2016,Murty2008}.

The objective of this paper is to present an experimentally-validated technical foundation for Intermediate-Energy Proton Irradiation (IEPI): a new materials irradiation technique that enables rapid, direct measurement of engineering properties with fidelity to advanced fission and fusion energy systems. The work presented includes new analytical tools for selecting proton energy to closely match the application environment, novel calculation of fundamental limitations including maximum dose-rates and thicknesses, and the first experimental demonstration that intermediate energy proton irradiation can predict bulk neutron-induced property changes. Utilizing this foundation, this technique can finely characterize the evolution of engineering material properties as a function of radiation damage, a critical need for cost-effective and safe design of nuclear systems, but one that is presently limited by the challenges of existing irradiation techniques.  Overall, the technique offers designers of these systems with a far more rapid and cost-effective tool for the evaluation, down selection, and qualification of materials than presently exists.  With private industry pursuing aggressive timelines both for new fission and fusion energy systems to be placed on the grid, the methods proposed here offer the potential paradigm shift in materials design and optimization necessary to realize these goals~\cite{Kuang2018,Ingersoll2014,Hejzlar2013}.

The paper is structured as follows: Section~\ref{Sec:Background} provides a short background on existing irradiation techniques -- irradiation in nuclear reactors or with micro-scale ion beams -- as context for understanding the advantages of of irradiation with 10+ MeV protons; Section~\ref{Sec:PKA} describes the fidelity with which protons can represent the recoil energies and damage cascades expected in materials during fission or fusion reactor operation; Section~\ref{Sec:Trans} simulates the ability of protons to represent the levels of helium and hydrogen production expected in fission and fusion materials; Section~\ref{Sec:temp} examines the relationship between the achievable dose-rates with proton irradiation, the degree of temperature uniformity, and the irradiated sample thickness; Section~\ref{Sec:Activation} investigates the radioactive hazard produced by intermediate-energy protons and compares the irradiation and dwell times to neutron irradiation; Section~\ref{Sec:Validate} presents the first irradiations and tensile tests of a bulk structural material with intermediate-energy protons, validating the proposed advantages of the technique; Section~\ref{sec:Impact} discusses the the results in detail and presents the impact on nuclear design.

\section{Background: evaluation of materials irradiation methods}
\label{Sec:Background}

This section provides a short summary and analysis of the advantages and limitations of the two most commonly used irradiation techniques: neutron irradiation in the core of nuclear reactors; and micro-scale ion irradiation with ion accelerators.  The section then introduces intermediate energy proton irradiation to show how this technique combines the key advantages of both techniques with several unique advantages.

\subsection{Nuclear reactor irradiation}
Radiation damage in materials has long been studied with neutron irradiation in fission test reactors. Due to the long mean-free path ($\sim$ cm) of reactor-spectrum neutrons in materials, the damage induced is inherently bulk, and the induced changes can be assessed in two complementary ways: first, through the use of macro-scale techniques such as tensile, fracture, and impact testing to directly extract engineering-scale properties; and second, through the use of techniques like electron microscopy, atom-probe tomography, and x-ray diffraction, that provide physical insight into the underlying microstructural evolution.

\begin{figure*}[pos=H!]
    \centering
    \includegraphics[width=0.75\textwidth]{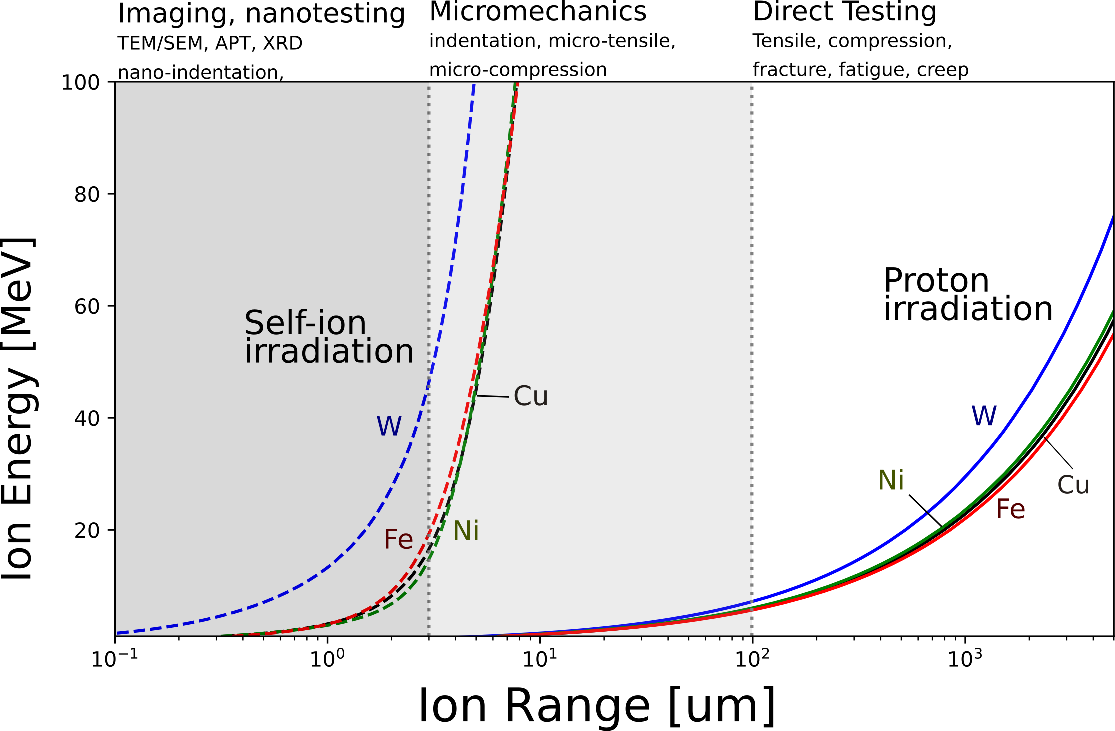}
    \caption{Ion ranges in four common elements demonstrate the ability of intermediate-energy protons to to achieve bulk penetration, allowing evaluation of irradiated material properties at the macroscopic engineering scale. Data from SRIM~\cite{Ziegler2010}.}
    \label{fig:Ranges}
\end{figure*}

Despite the advantages of reactor irradiation, its high costs and time requirements limit its use as a tool for designing new nuclear materials and components. Because advanced fission and fusion energy system designs can have material dose requirements approaching 200 DPA, thermal test reactors ($\sim$5 DPA/year) and even fast test reactors ($\sim$20 DPA/year) are not practical options for frequent high dose experiments~\cite{CNM_ions,Zinkle2013}. Irradiation in test reactors typically leads to high levels of radioactivity, requiring significant post-irradiation dwell times (months -- years) and specialized safety equipment during post-irradiation analysis. Due to the complexities of operating fission reactors, test reactor experiments can cost millions of dollars and can require years or even a decade to complete~\cite{CNM_ions}.  These limitations of reactor irradiation manifest as extremely long design and development cycles for new nuclear materials; for example it is expected that to 20-25 years are needed just to design and implement an improved nuclear fuel and cladding material in already existing nuclear reactors~\cite{Barrett2012}. While other advanced materials applications benefit from iterative cycles of design and experimentation, the extreme time and cost limitations prevent such iteration and improvement in nuclear systems. 

The difference in neutron energy spectra between current test reactors and future (e.g. fusion) reactors is a further obstacle to the design of reactor components and materials. For example, a tungsten component exposed to neutrons in a fusion reactor is expected to produce several atomic-parts-per-million (appm) of helium per year~\cite{Gilbert2012,Gilbert2015c}, which will stabilize voids and promote swelling~\cite{Was_book}. Both thermal and fast reactors lack the high energy neutrons required to generate comparable amounts of helium~\cite{Gilbert2015,Gilbert2015a,Gilbert2015b}. Conversely, the large fraction of thermal neutrons in test reactors, which are not present in fusion first wall materials, cause artificially high production of dissolved impurities and the formation of embrittling precipitates~\cite{Katoh2019}. These inherent differences in neutron energy spectra limit the fidelity with which future fusion and fission reactors can be designed based on testing with current test reactor technology.

\subsection{Micro-scale ion irradiation}
The irradiation of materials with ion beams has been widely adopted by the nuclear materials community as a faster, more flexible tool for studying radiation damage effects and designing higher performance nuclear materials~\cite{CNM_ions}. Large ion currents (\SI{1}{\micro\ampere} to 1 mA) enables dose rates on the order of DPA/day-DPA/h~\cite{Was2002} providing access to the life-time doses expected in advanced fission and fusion systems in practical experimental times. Most ion irradiations are performed with relatively low energy ($\sim$ MeV) ions and avoid any radiation-producing nuclear reactions. This allows relatively low-cost and flexible experimental facilities and accelerates post-irradiation testing. While neutron irradiation has remained the gold-standard for assessing materials, proton irradiation has reproduced the material response of neutron irradiation across a variety of microstructural effects (e.g. segregation and loop formation) and property changes (e.g. hardening)~\cite{Was2002}. 

The use of most ion irradiation techniques for rapid the design of nuclear systems is limited by the microscopic thickness of the induced damaged. Due to technology and cost limitations, the vast majority of ion accelerators available for materials irradiation provide relatively low energy light ions (e.g. protons $\lesssim$5 MeV) or high Z ions (e.g. self-ions, swift heavy ions)~\cite{IAEA_database}. The ranges of such beams are typically micrometers or less, with sharp spatial gradients in damage levels. This range limitation complicates the ability to extract engineering properties, restricting applicable tests to micro- or nano-scale techniques such as indentation testing and preventing bulk measurements of strength, ductility, toughness, and creep. The limited micro-scale testing available to ion-irradiation experiments is susceptible to distortion by the effects of denuded zones~\cite{Zinkle2018}, injected ions~\cite{Zinkle2018}, and carbon contamination~\cite{Shao2019,Was2017}, which are purely artifacts of ion irradiation. Polycrystalline, multi-phase, or composite materials are difficult or impossible to study with conventional ion techniques because the material length scales of interest are much larger than the irradiation regions.  Therefore, the use of conventional ion irradiation techniques can only provide limited information to aid the design and qualification of new nuclear materials.

\subsection{Intermediate energy proton irradiation}
Intermediate-energy (10+ MeV) protons offer the possibility of combining the benefits of ion irradiation (e.g. high dose rate, flexible irradiation conditions) and reactor neutron irradiation (uniform/bulk irradiation, direct engineering testing). As the lightest ion, protons have the lowest stopping power and greatest range in materials. As shown in Figure~\ref{fig:Ranges}, protons above 10 MeV are able to penetrate more than \SI{100}{\micro\meter} in typical metals - a length scale at which direct testing of mechanical properties becomes feasible - without inducing the the confounding effects associated with low-range ions.  As with all ion irradiation, high dose rates are achievable and can be enhanced by increasing the ion currents. Unfortunately, accelerators providing protons with energy above 10~MeV (and beam currents above $\sim$\SI{1}{\micro\ampere}) have been rare due to their high capital and operating costs, large physical size, electricity consumption, and staffing requirements. As a result, these machines are typically too expensive for widespread materials development research and are dedicated almost exclusively to nuclear and astrophysical research~\cite{Gupta2020} or the production of isotopes for research and medicine~\cite{Review_MedIsotope}.

Recent accelerator technology advances have opened the possibility of widespread use of intermediate energy protons to support nuclear material design. Commercially available, ultra-compact, superconducting cyclotrons are now able to produce intermediate-energy proton beams of relatively high currents (10 -- 100~uA)~\cite{Vincent2016}. Utilizing this technology, a new university-scale facility has been dedicated to materials irradiation research to support fusion and advanced fission reactor design~\cite{Jepeal}.  In order to effectively use this and other existing and future facilities to accelerate the pace of nuclear materials design, there is a need to understand the opportunities and fundamental limitations of intermediate energy proton irradiation in predicting reactor-relevant radiation damage. The following sections explore these aspects of intermediate energy proton irradiation in detail.

\section{Irradiation fidelity through recoil energy spectra matching}
\label{Sec:PKA}
In a collision between an irradiating particle and a primary knock-on atom (pka), the amount of energy transferred can vary over many orders of magnitude, from eV to MeV, leading to drastically different damage cascades and subsequent primary damage production~\cite{Was_Displacement}.  The inherent difference between charged particle interactions and neutron interactions drives different amounts of energy transfer to pkas. For example, 1 MeV protons in copper have a much lower average recoil energy (500 eV) than 1 MeV neutrons (45 keV). Likewise, the proton recoils are evenly distributed over three orders of magnitude in pka energy (20~eV -- 50~keV), while over 95\% the neutron recoils fall within a single order of magnitude (5 -- 50~keV)~\cite{Averback1994}. This difference in recoil energies is often used to argue for the benefit of heavy ions, which have a higher average recoil energy than protons (but have an similarly wide distribution in energies).  However, in order to make an accurate comparison to true reactor conditions, charged particle pka energies must be compared to a reactor spectrum of neutron energies, rather than neutrons of only a single energy.

\subsection{Generating weighted recoil spectra}
\label{subsec:PKA spectrum}
In order to evaluate the similarity of proton, neutron, and other ion irradiation techniques to realistic conditions, a case study of recoil energies was performed using tungsten as an example of a fusion-relevant material. Recoil energy distributions are calculated for mono-energetic proton and self-ion exposure, along with a fast fission reactor neutron spectrum and a hypothetical fusion reactor neutron spectrum. These recoil spectra are weighted according to a common energy-based weighting and compared graphically.

The initial recoil spectra are each taken from previously published work and given a consistent weighting. Ion-induced recoil energy spectra were produced from the code DART, which is informed by binary collision approximations~\cite{Luneville2006}. The fast reactor recoil energy spectra was taken from the FISPACT Materials Handbooks for fast reactors~\cite{Gilbert2015b} using the simulation of in-core conditions.  Likewise the fusion reactor recoil spectrum was taken from the FISPACT Handbook for fusion reactors~\cite{Gilbert2015c}, representing a fusion first-wall component.  Spectra are weighted according to the NRT formula~\cite{Norgett1975}, which predicts the number of defects created by each recoil based on the recoil energy. The FISPACT software assumes a tungsten displacement energy of 55~eV~\cite{SUBLET2017}, and this value was also used in all other calculations. 

\begin{figure}
    \centering
    \includegraphics[width=0.48\textwidth]{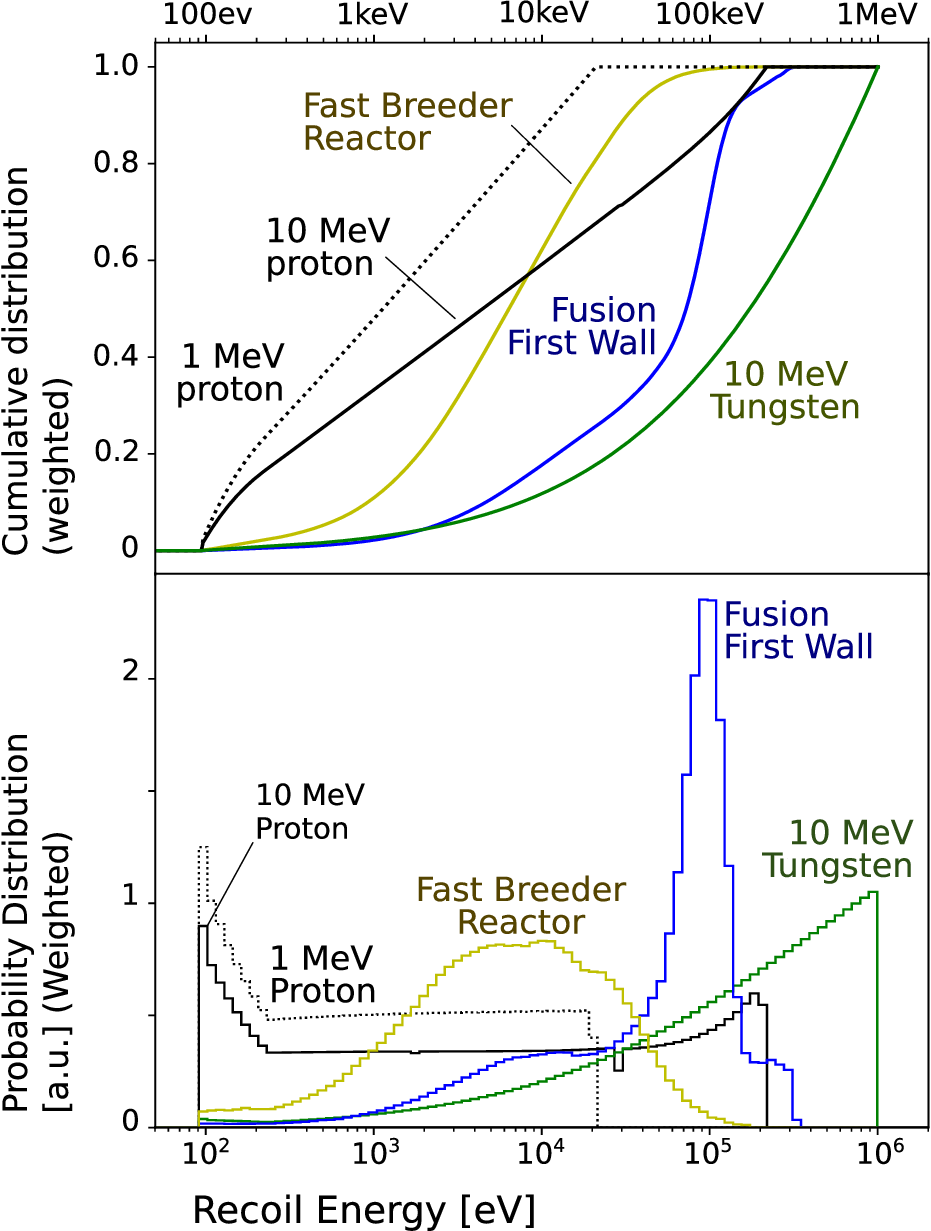}
    \caption{The simulated pka energy spectra under example reactor neutron spectra and representative monenergetic ions. Shown here are the weighted cumulative distribution function (top) and weighted probability distribution histograms (bottom) of tungsten atoms calculated with the DART code~\cite{Luneville2006} or FISPACT code~\cite{Gilbert2015c,Gilbert2015b}.} 
    \label{fig:RecoilEnergy}
\end{figure}

Several weighted recoil spectra for tungsten are plotted in figure~\ref{fig:RecoilEnergy}, both as cumulative distribution functions (CDF) and as probability distribution functions (PDF).  The CDF and PDF are two equivalent ways for presenting the spectrum of pka energies that are produced by each irradiating particle; While the CDF is a more common way to present pka energy ranges, the PDF allows more easy comparison of the similarity between irradiation types, which visually represented as the overlap between the two spectra.

Figure~\ref{fig:RecoilEnergy}, demonstrates that both the fusion neutron spectrum and the fast breeder spectrum lead to pka energy distributions spanning several orders of magnitude, which contrast with the narrower distribution expected from monoenergetic neutrons. Due to the wide spread in neutron recoil energies, intermediate-energy protons have similar recoil spectra to the fusion and fast fission reactor cases. Intermediate-energy protons are able to create recoils fairly uniformly distributed across the range of recoils expected in fission and fusion reactors (up to 100 eV -- 300 keV). Lower energy protons ($\sim$1 MeV) are only able to create low energy recoils (up to 10s of keV). Likewise, 10 MeV tungsten self-ions drastically over-produce very high energy recoils (up to MeVs) which are not expected from fusion or fission reactors.  An approximate similarity between the recoil spectrum can be seen from figure~\ref{fig:RecoilEnergy}, but there is still a need to systematically evaluate this similarity over a wide range of ion energies.

\subsection{Optimizing the ion energy for recoil similarity}
\begin{figure*}
    \includegraphics[width=00.96\textwidth]{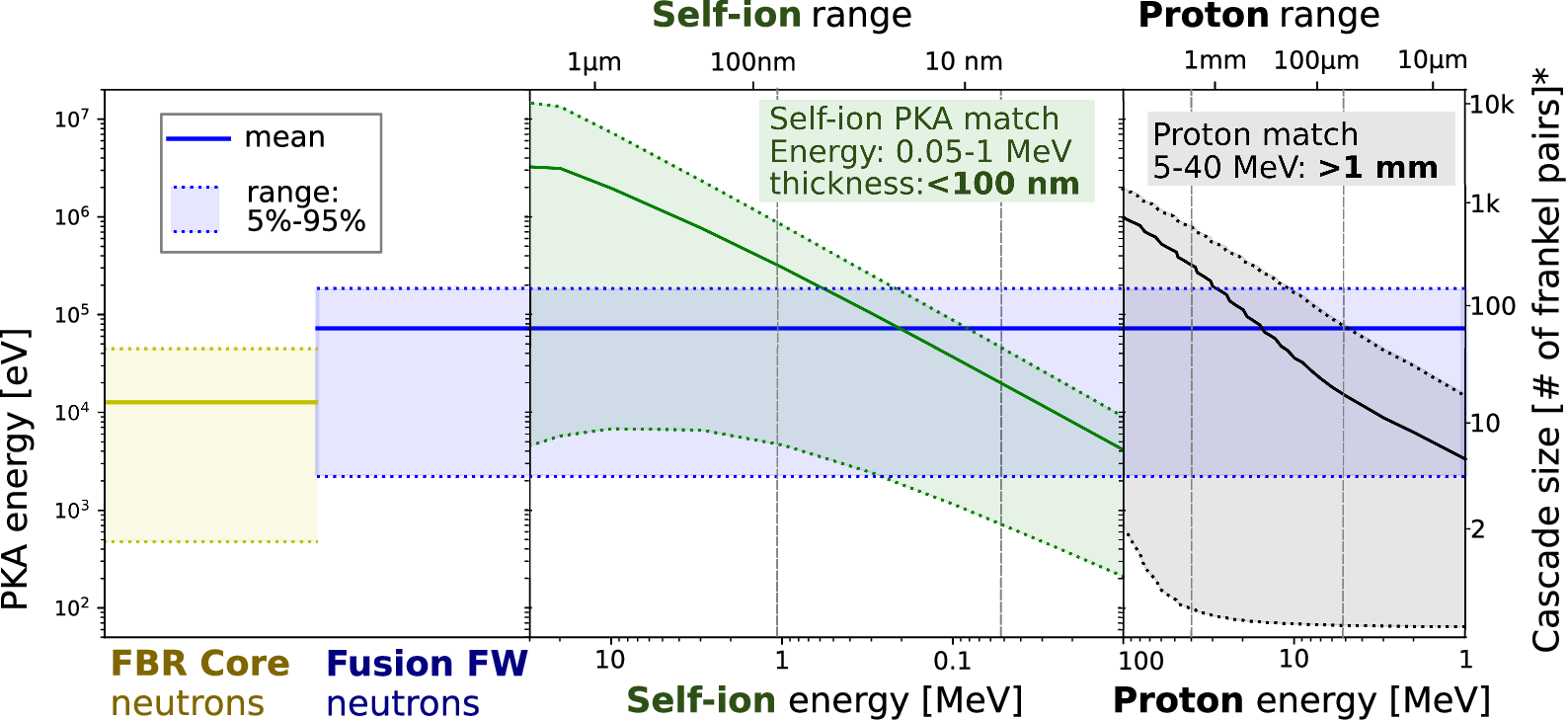}
    \caption{Comparison of tungsten recoil energies from four different irradiation scenarios. Fixed neutron spectra in a fast breeder reactor (FBR) core are not well matched; self-ion and proton kinetic energies can be tuned to provided the greatest fidelity to the fusion first wall case. Note, however, that optimized proton irradiation fidelity occurs with ranges suitable for bulk specimens whereas self-ion irradiation is limited to nanoscale depths.}
    *Frenkel pairs calculated according to the ARC-DPA formulation~\cite{Nordlund2018}
    \label{fig:RecoilSim}
\end{figure*}

The recoil energy spectrum for a given ion species depends strongly on its incident kinetic energy, as demonstrated by the difference between the 1 MeV and 10 MeV proton curves in figure~\ref{fig:RecoilEnergy}. Thus, an ion irradiation experiment can be tailored to better match the expected material response in the real world application by carefully selecting the incident ion kinetic energy. In order to make a more general comparison between protons and heavy or self-ions it is beneficial to establish metrics for recoil spectrum similarity that can be compared across many potential irradiation scenarios. The weighted average recoil energy for each irradiation scenario was calculated by:

\begin{equation}
    \label{Eq:av}
    \overline{E}_{pka}=\int _0^{\infty}E*P(E)*dE
\end{equation}

where $\overline{E}_{pka}$ is the weighted average energy, $E$ is the pka energy, and $P$ is the weighted probability distribution function. Along with the weighted average energy, a spectrum range was calculated as the energy bounds within the middle 90\% of recoils fall. Both the weighted average and the energy ranges are plotted in figure~\ref{fig:RecoilSim} for the tungsten-fusion case. As with section~\ref{subsec:PKA spectrum}, the neutron pka spectra~\cite{Gilbert2015b,Gilbert2015c} and ion pka spectrum~\cite{Luneville2006} are taken from previously published work and weighted by the NRT formula~\cite{Norgett1975}.

Figure~\ref{fig:RecoilSim} demonstrates an intrinsic and limiting trade off in self ion irradiation as well as the lack of fidelity in reactor irradiations. At low irradiation energies (< 1 MeV), the recoil spectra of tungsten in self-ion irradiation provide a close match to the material response expected for tungsten in the fusion environment; however, the range of tungsten ions at 1~MeV is less then \SI{100}{\nano\meter}, resulting in challenging irradiation and post-irradiation measurement constraints and complicating the extrapolation of results to larger length scales of interest in engineering applications. To combat this, tungsten self-ion experiments are routinely performed at higher energies, such as 20~MeV~\cite{Ogorodnikova2015, Armstrong2013, Xu2015, Hwang2016} to provide radiation uniformity over larger sample depths. At such high ion irradition energies, the weighted average pka energy is more than an order of magnitude higher than in the fusion environment, inducing cascades of many thousands of displacements with high energy self-ion irradiation compared to the fusion application with cacades of tens to hundreds. A similar problem in matching the pka response of tungsten in a fusion environment exists for emulating fission reactors, although in this case the recoil energy spectra is lower by an order of magnitude as shown by the fast breeder reactor case in figure~\ref{fig:RecoilSim}). In contrast, proton irradiation with 5 -- 40~MeV simultaneously achieves both high fidelity matching to the pka response of tungsten in a fusion environment with uniform irradiation depths exceeding  1~mm, over four orders of magnitude larger than self-ion irradiation. This enables straightforward bulk irradiation of macroscale specimens suitable for direct engineering assessments like tensile tests with high confidence in the fidelity of the material response to a fusion environment.

\subsection{Discussion}
This case-study demonstrates that intermediate-energy protons can have a high degree of similarity in recoil energies to future reactor conditions. Protons produce recoils with a wide distribution in energy, and the range of that distribution can be tailored by changing the proton energy. Similarly, realistic reactor conditions also produce a wide distribution in recoil energies, with a comparable energy range and mean pka energy to intermediate-energy protons.  Other ions, such as self-ions, can also produce tailored recoil spectra, but the desired recoil spectra come from lower ion energies and leading to much more limited range. While self-ion experiments are inherently a compromise between increasing the ion range and maintaining a representative recoil energy spectrum, proton irradiation allows both extended range and recoil similarity to occur simultaneously.

While this case-study focused on tungsten as a fusion plasma-facing material, the methodology is extensible to any hypothetical fusion power plant or advanced nuclear reactor component by substituting the specific neutron spectrum and material of interest.  Therefore, this analysis not only indicates characteristics of proton and self-ion irradiation, it also established a framework for quantitative comparison of recoil similarity between many different irradiation scenarios.

\section{Proton induced transmutation: fidelity to nuclear environments}
\label{Sec:Trans}
In addition to displacing atoms, neutrons from fission and fusion reactors also cause damage by introducing impurity atoms into the irradiated materials. Neutrons are capable of generating nuclear reactions, including ($n,\alpha$) and ($n,p$) reactions, which transmute the original atoms into new atoms, often including low levels of helium and hydrogen, in the range of atomic parts per million (appm). Appm concentrations of helium are well known to exacerbate radiation damage effects, promoting the formation of voids, causing high temperature helium embrittlement, and weakening grain boundaries at low temperatures ~\cite{CNM_Helium}.  Additionally, hydrogen has been shown to further promote the formation of voids through synergistic interaction with helium~\cite{Marian2015}.  The combined effect of helium and hydrogen on radiation damage is not fully understood, and are the study of ongoing research using highly specialized techniques such as multiple-simultaneous-ion-beam experiments with coincident heavy ions, helium ions, and protons~\cite{Marian2015}. While these existing techniques allow insight into He and H damage effects, they produce damage that is shallow, nonuniform, and therefore not representative of bulk neutron damage~\cite{CNM_Helium}. 

Intermediate energy protons also possess the ability to generate transmutation products including helium and hydrogen simultaneously with displacement.  Common nuclear materials have the potential for nuclear reactions such as ($p,\alpha$) and ($p,p'$), resulting in the helium and hydrogen generation analogous to neutron reactions.  Cross sections for proton induced nuclear reactions vary with proton energy; therefore. there is the potential for the amount of transmutation produced to be controlled by modifying the incident proton energy. In order to evaluate if intermediate-energy protons could serve as a source of radiation damage with the reactor-relevant levels of transmutation, there is a need to systematically compare proton and neutron induced transmutations.

In this section, a comparison of transmutation products is made between fission reactor, fusion reactor, and intermediate-protons. Transmutation calculations normalized across irradiation types by the number of displacements and were performed on common base elements for nuclear materials. A wide range of proton energies were used to evaluate the degree to which proton irradiation can be tailored to match a given neutron environment.

\subsection{Predicting proton-induced transmutation}

The proton-induced transmutation was calculated using the inventory and activation code FISPACT~\cite{SUBLET2017}.  In these calculations, the proton irradiation was modeled as mono-energetic protons with a consistent flux of $6.24\times10^{14}~\text{cm}^{-2}\text{s}^{-1}$ and irradiation time of 1 day (a representative flux and irradiation time for intermediate-energy proton experiments~\cite{Jepeal}). Helium and hydrogen levels were extracted at the end of the simulated irradiation, yielding the average amount of transmutation per unit of proton fluence.  The damage efficiency (number of DPA per proton fluence) was calculated at each proton energy using the DART program, assuming the same displacement energies as the FISPACT software (55~eV for tungsten, 40~eV for copper, nickel, and iron)\cite{SUBLET2017}. Using this transmutation yield and damage efficiency data, the transmutation was normalized to displacements  (appm He and H per DPA) each proton energy and element.

The neutron-induced transmutations performed with FISPACT using example reactor neutron spectrum provided in the FISPACT Handbooks~\cite{Gilbert2015a,Gilbert2015b,Gilbert2015c}. These simulations were performed up to 1 DPA, and the helium and hydrogen impurity levels were recorded at the end of the simulated irradiation. These simulations were checked against published simulation data in the FISPACT handbooks to ensure their validity.

\subsection{Comparison to nuclear environments}

\begin{figure*}[pos=t!]
        \centering
    \includegraphics[width=0.95\textwidth]{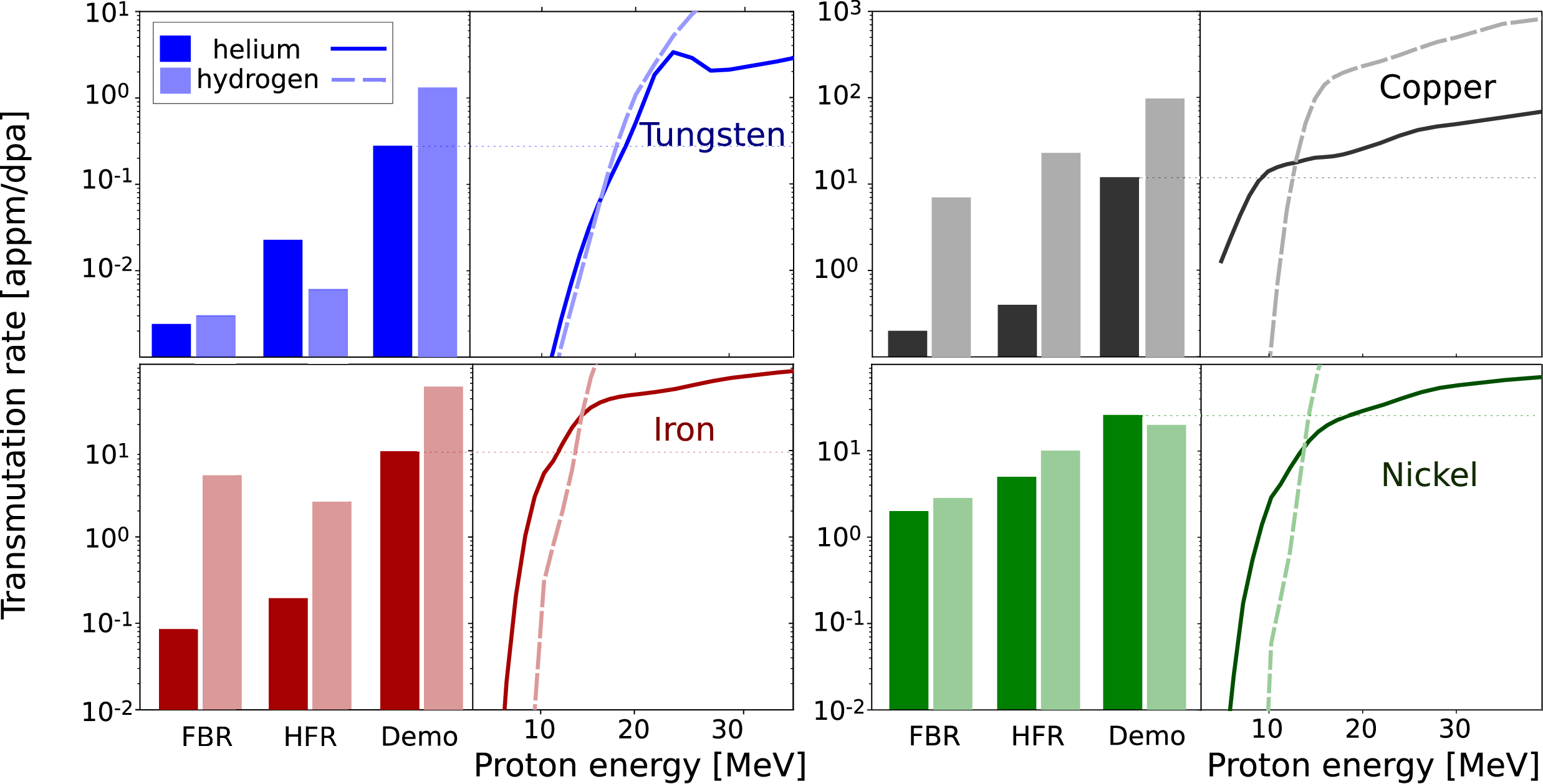}
        \caption{Transmutation calculations demonstrate that intermediate energy protons can generate fission and fusion relevant levels of helium and hydrogen during irradiation.}
        \label{fig:Trans}-
\end{figure*}

In figure~\ref{fig:Trans}, the normalized production of helium and hydrogen in materials are compared for three neutron environments (fast breeder reactors (FBR), high flux reactors (HFR) and prototypical fusion power plant first wall (Demo)) and proton irradiation. There are three important takeaways from this figure. First, there are orders of magnitude difference in the production of helium and hydrogen across neutron environments with a fusion power plant typically producing one to two orders of magnitude more than either fission reactor environment due to the presence of 14.1 MeV neutrons from the deuterium-tritium fusion reaction. Material irradiation in fission reactor are insufficient to reproduce the important effect of H and He accumulation found in fusion materials; any high-fidelity irradiation technique must be able to produce H and He over the range required to achieve fusion relevant levels of these transmutation gases.

The second takeaway is that - due to the nuclear reaction thresholds that generate H and He in charged particle reactions with the irradiated material - low energy proton irradiation (below 5-10 MeV) and self-ion irradiation for materials such as tungsten cannot generate transmutation gases that important to capturing the full material response of fusion materials.

The final and most important takeaway is the capability of intermediate energy protons to produce fusion-relevant quantities of H and He transmutation gases over three to four order of magnitude in concentration simply by tuning the irradiation energy. For each element, the amount of helium or hydrogen produced could be matched to the application with a proton energy of less than 20~MeV. This matches closely with the energy range required to achieve high fidelity in the pka spectrum recoil distribution and irradiation of bulk specimens suitable for direct engineering testing discussed in Section~\ref{Sec:PKA}. Because the helium and hydrogen generation are coupled, it is difficult to simultaneously match both the helium and hydrogen production to the application. However, proton irradiation does allow, in general, a closer match to a fusion environment than either of the fission reactor environments, which produce much less He and H in each case. Additionally proton irradiation at several energies would produce comparable irradiations with a varied level of He and H production, allowing a controlled study of the impact of helium and hydrogen production. Therefore, intermediate-energy proton irradiation can be tailored to study the transmutation effects expected in fusion and future fission reactors with a much greater range than multiple-beam techniques and much more flexibility than fission irradiation.

\section{Predicting proton irradiation constraints: dose rate, sample thickness, and temperature uniformity}
\label{Sec:temp}
Because protons directly heat their samples during irradiation, the dose-rate, sample thickness, and temperature uniformity are inherently coupled for a proton irradiation experiment.  While this beam-heating phenomena is well known, there is a lack information about the inherent dose-rate limits set by heating.  To address this knowledge gap, modeling of interaction of protons with sample materials was performed to understand what experimental conditions are achievable. This modeling captures the deposition of heat into a sample by a 12 MeV proton beam, the conduction of that heat under ideal circumstances, and the amount of damage created by the protons. This establishes bounds on what dose-rates and sample thicknesses are achievable for 12 MeV proton irradiation, as a representative case for all intermediate-energy ion irradiation. A 12 MeV beam was used for this analysis because it is the beam energy available to the materials irradiation facility that performed the irradiation demonstration in section \ref{Sec:Validate}~\cite{Jepeal}; however, the methodology and modeling tools developed for this assessment can be easily applied to any irradiation energy.

\subsection{Modeling of heat and damage production by protons}

The first step in this analysis was to model the heat deposition by the proton beam into a sample.  Range vs. Energy data was extracted from SRIM tables~\cite{Ziegler2010} to model the proton's energy loss through the sample, and was interpolated to create a table of proton energy values as a function of depth (beginning with an incident proton energy of 12~MeV). All energy lost by the beam was assumed to translate to heat production locally in the material, ignoring the small fraction that remains as defects.

\begin{figure}
        \centering
        \includegraphics[width=0.48\textwidth]{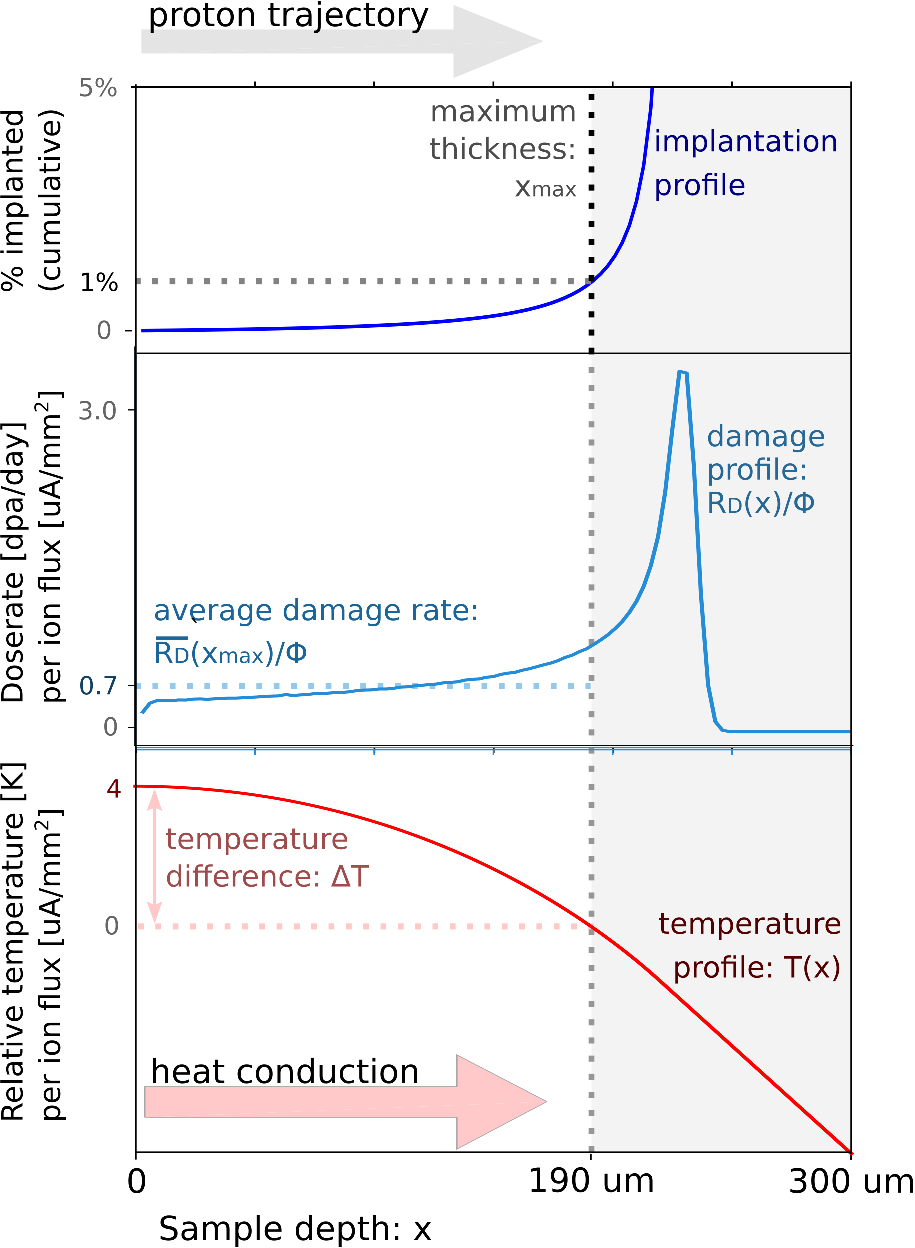}
        \caption{Example modeling of proton-tungsten interactions demonstrating how results were calculated for figure~\ref{fig:Temperature}. Profiles were calculated from TRIM simulations and SRIM range tables~\cite{Ziegler2010}}
        \label{fig:profiles}
\end{figure}

With the knowledge of heat production, the heat transfer is then modeled.  Heat conduction is assumed to be purely 1-dimensional - the ideal for minimizing temperature differences in a sample - and heat is assumed to conduct along the direction of the incident protons. The increase in temperature is tracked through the sample thickness, using the room temperature thermal conductivity of the material.  With these assumptions, the temperature difference from one end of a sample to the other can be calculated by:

\begin{equation}
    \begin{split}
    \label{Eq:T}
        \Delta T &= T(x_{max})-T(0)=\int_{0}^{x_{max}} \frac{dT}{dx}(x) dx = \int_{0}^{x_{max}} \frac{q(x)}{k}dx \\ &= \frac{\Phi}{k} \int_{0}^{x_{max}}(E(x) - E(0)) dx
    \end{split}{}
\end{equation}{}

\noindent where $\Delta T$ is the total temperature difference across the sample, $T(x)$ is the temperature at a depth $x$, $E(x)$ is the proton energy, $\Phi$ is the incident flux of protons, $k$ is the thermal conductivity, and $q(x)$ is the heat flux. A representative temperature profile is shown in figure~\ref{fig:profiles}, demonstrating how $\Delta T$ is calculated. 

With a given $k$, $E(x)$, and $x_{max}$, Equation~\ref{Eq:T} yields the maximum flux allowed for a specific temperature variation in a sample. This method was used in this analysis to calculate the maximum flux for a 5~K temperature difference across a range of thicknesses (0 -- \SI{400}{\micro\meter}) for four pure metal elements (Fe, Cu, W, Ni).  This conservative limit of 5~K was set as an optimistic case for temperature uniformity, knowing that any real experimental system will have greater difference due to imperfect conduction, and as a representative temperature difference that could have some impact on defect mobility.  Because equation~\ref{Eq:T} presents a linear relationship between temperature difference and proton flux, these results can be scaled linearly to any other desired temperature limit. 

\begin{figure}[pos=b]
        \centering
        \includegraphics[width=0.48\textwidth]{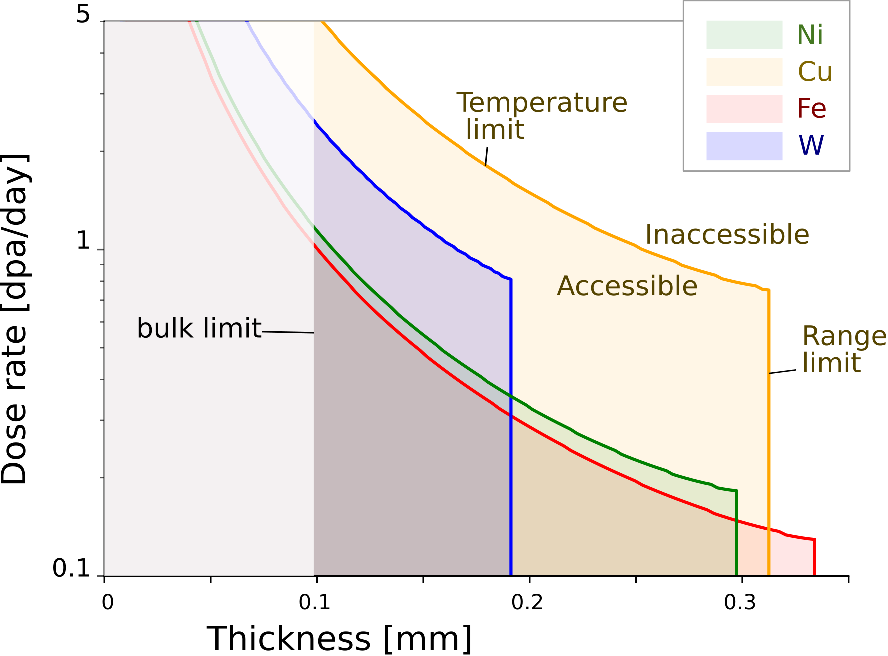}
        \caption{Accessible regimes for 12~MeV proton irradiation including a temperature uniformity requirement (max 5~K difference) and avoiding proton implantation (<1\% of protons implanted)}
        \label{fig:Temperature}
\end{figure}

Using TRIM simulations, the maximum allowable flux values were converted to maximum allowable dose-rates. For each material, a TRIM simulation was constructed simulating 12 MeV protons, as shown in figure~\ref{fig:profiles}. Each TRIM simulation output the density of vacancies created, per unit of ion-fluence, at each depth location into the sample. This damage profile was converted into an average dose rate per unit flux by equation~\ref{Eq:DPA_rate}:

\begin{equation}
    \begin{split}
    \label{Eq:DPA_rate}
   \frac {\overline{R_D}(x_{max})}{\Phi}&=\frac{1}{x_{max}}\int_0^{x_{max}}\frac{R_D(x)}{\Phi}dx \\
    &= \frac{1}{x_{max}*\rho_N}\int_0^{x_{max}}\frac{\rho_V}{\Psi}(x)dx  
    \end{split}{}
\end{equation}

\noindent where $\frac{R_D}{\Phi}$ is the displacment rate [DPA/s] per proton flux [p/cm\textsuperscript{2}$\cdot$s], $\rho_N$ is the atomic density of the target [atoms/A$^3$], and $\frac{\rho_V}{\Psi}$ is the vacancy density output from SRIM [$\frac{vac}{A^3}/\frac{ion}{A^2}$]. This value for $\frac{\overline{R_D}}{\Phi}$ from equation~\ref{Eq:DPA_rate} is then multiplied by the maximum flux calculated from equation~\ref{Eq:T}, yielding the maximum doserate achievable for a given temperature difference and material scenario. 

Maximum dose-rates for the four materials and a range of thicknesses are plotted in figure~\ref{fig:Temperature}. A range limit is calculated directly from the TRIM simulations as the maximum distance that 99\% or more of the protons travel through the sample rather than implanting into the sample (example shown in figure~\ref{fig:profiles}). A "bulk limit" is also indicated at 100um, as the approximate sample thickness for which bulk mechanical properties can be easily tested through direct means such as tensile testing\cite{Kumar2016,Takeda2017}. 

\subsection{Dose-rates and thickness bounds for representative metals}
\label{Subsec:temp_results}

As shown in figure~\ref{fig:Temperature}, for each example material, 12~MeV proton irradiation is able to access a range of thicknesses greater than 100~\micro m and dose-rates on the order of 1~DPA/day without causing temperature differences greater than 5~K through the thickness of the sample. The exact shape of the accessible area is dependent on many material properties, including thermal conductivity, displacement threshold energy, and the stopping power of protons in the material (which depends on density, atomic number, etc.). Despite these differences, each material allows irradiation to rates greater than 0.1~DPA/day in relatively thick samples (200 -- 300 \micro m) and irradiation to rates as high as  $\sim$1 DPA/day in thinner samples ($\sim$100 \micro m).  Furthermore, irradiating any of the example materials to dose-rates in excess of 10~DPA/day would require either very thin samples (50 -- 100 \micro m) and/or large temperature inhomogeneity in the sample (in addition to a $\sim$mA source of intermediate-energy protons). This analysis demonstrates that the range of 0.1 -- 1~DPA/day with samples thicknesses of 100 -- 300 \micro m is readily accessible to proton irradiation without compromising temperature homogeneity.

\subsection{Discussion}

The results of this case-study of 12 MeV protons presents the basic trade-off between achievable experimental conditions for all intermediate-energy protons. While extended range is desired for easier measurement of bulk properties after irradiation, it comes with an approximately quadratic reduction in the dose-rate achievable (e.g. doubling range  decreases doserate to one-fourth). Increasing the incident proton energy is an option to further increase range, but that increased range will further decrease the achievable dose-rates, which is likely the limiting factor for the speed of experiments.  Instead, many proton experiments will benefit from designing for the thinnest samples that can be easily measured, maximizing the dose-rate and temperature uniformity achievable.

Additionally, this work also presents a novel framework for understanding the limitations of proton (or other light-ion) irradiation under any arbitrary scenario. With modification of the inputs to TRIM and SRIM and to the thermal conductivity, this framework could be adapted to any material, proton energy, and irradiation temperature. As such, it is an flexible method for predicting the fundamental limits of irradiation experiments, due to the inherent characteristics of the ion and materials chosen. Therefore, this framework can help determine the fundamental limitations of intermediate-energy proton irradiation in any future scenario. 

\section{Predicting and reducing irradiation-induced radioactivity}
\label{Sec:Activation}

Both neutron and intermediate-energy proton irradiation experiments can cause radioactivity in irradiated materials and risk to personal and environmental safety. Neutrons and protons (above an isotope-dependent energy threshold $\sim$ 1-10 MeV) lead to formation of new isotopes through the process of transmutation. The delayed release of radiation through the decay of long-lived radioisotopes poses a prolonged obstacle to the safe evaluation of irradiated samples.  Addressing the hazard of radioactive samples often requires long times (months - years) to allow radiosotopes to decay, expensive equipment (e.g. gloveboxes, dosimeters) to protect personnel, and limits examination of the material response (requiring remote handling and limits to time spent near samples).  Mitigating this radioactive hazard is paramount to allowing frequent and cost-effective measurements of radiation damage.

This section compares intermediate-energy proton and reactor activation of irradiated materials in order to understand the capability for protons to enable low-activation, bulk irradiation experiments.  Predicting proton and neutron activation of materials is complex, requiring precise knowledge of the incident radiation energy spectra, energy-dependent cross-sections for each isotope present, and half-lives of each radioisotope produced. As a result of strong variations in each of these quantities, it is challenging to make general claims about the levels of radioactivity across different reactor or proton irradiation scenarios. In order to make a direct and representative comparison of reactor and intermediate-energy proton irradiation, a case-study of induced radioactivity was performed across a set of four engineering materials.  In this study, the irradiation-induced activation of each material was calculated for each irradiation scenario: 12 MeV proton, high flux reactor, and fast reactor exposure, each to an equivalent level of exposure. The decay of this activation over time after the irradiation was predicted to quantify both the level of hazard and the potential time-cost across techniques. 

\subsection{Simulating activity after irradiation}

Both neutron and proton simulations were performed using the FISPACT code~\cite{SUBLET2017}. These simulations require inputs including an irradiating particle energy spectrum, irradiation flux, irradiation time, and material composition. The simulations output a variety of data relating to transmutation, activation, and radiation exposure in the irradiated material. 

Representative values for neutron energy spectra and fluxes were taken from the FISPACT materials handbooks for fast breeder reactors~\cite{Gilbert2015b} and high flux reactors~\cite{Gilbert2015a}. Initial simulations ouput a doserate in DPA/day, which was used to calculate the time of irradiation needed to reach 1 DPA. This irradiation time to 1DPA was used as the time input for further simulations.

\begin{table*}[pos=p!]
    \centering
    \begin{tabular}{l|l|l|l}
Material   &Major      & Minor  & Impurities\\
name       &elements   & elements&\\
\hline
Tungsten ITER-grade & W 99.94           &
&C 0.01, O 0.01, N 0.01, Fe 0.01, Ni 0.01, Si 0.01 \\
&&&\\ 
CuCrCZ ITER-grade  & Cu 98.98      & CR 0.75, Zr 0.11   
&Nb 0.1, Co 0.05, Ta 0.01\\
&&&\\
Stainless steel 316*   & Fe 64.6, Cr 17.5  & Mo 2.5, Mn 1.8
& C 0.023, N 0.007, Ti 0.15, Si 0.5, Cu 0.3 \\
        ITER-grade&  Ni 12.3              & 
&Co 0.05, P 0.025, Nb 0.01, Al 0.15\\
&&&\\ 
Incoloy 908  & Ni 49.5, Fe 40.7 & Cr 3.9, Nb 3.0, 
& Si 0.15, Mn 0.04, C 0.01, Cu 0.01, Mo 0.02, Ta 0.01\\
        &                   & Ti 1.6, Al 1.0 
&  P 0.003, B 0.003, S 0.001, O 0.001, N 0.001\\
\multicolumn{4}{l}{}\\
\multicolumn{4}{l}{*other trace elements included, see Appendix}
    \end{tabular}
    \caption{Material compositions for the four fusion relevant materials used in the FISPACT activation simulations }
    \label{Tab:comp}
\end{table*}

Proton spectra and fluxes were set to be representative of a realistic proton irradiation experiment, as is described later in section~\ref{Sec:Validate}.  Proton spectra were calculated assuming a sample thickness of \SI{150}{\micro\meter}, which is shown in section~\ref{Sec:temp} to enable bulk property measurement while limiting temperature differences due to beam-heating. SRIM range tables were used to determine how much material was exposed to each proton energy as the proton beam degraded from its initial energy (12 MeV) to its exit energy ($\sim$ 3-5 MeV).  Proton fluxes were set to $6.24\times10^{14}~cm^{-2}s^{-1}$ to represent a beam current density of \SI{1}{\micro\ampere}/mm$^{2}$ (representative of values used in~\cite{Jepeal}).  Dose-rates were calculated from TRIM simulations using the methodology described by equation~\ref{Eq:DPA_rate}, and using the displacement energies specified in the FISPACT user manual\cite{SUBLET2017}. These dose-rates were then converted into time of irradiation needed to reach 1DPA which was used as an input to FISPACT.

\begin{figure*}[pos=p!]

   \includegraphics[width=.95\textwidth]{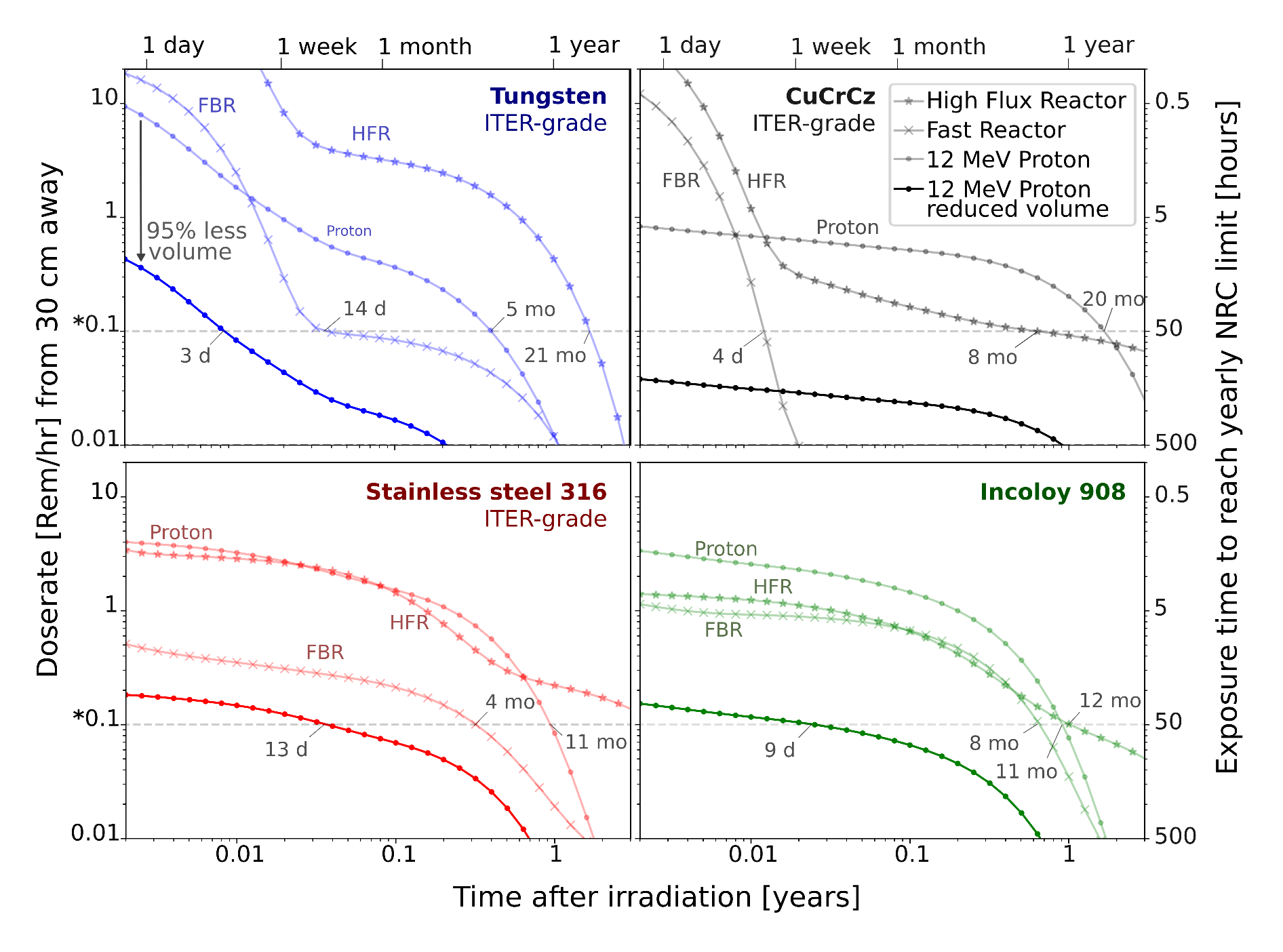}
    *NRC threshold for a "high radiation area"~\cite{NRC_doserate}
    \caption{Simulated radioactive hazard level for proton and neutron experiments demonstrates the advantage in dose to persons and/or cooldown time using proton irradiation}
    \label{fig:Activation}
\end{figure*}

\begin{figure*}[pos=b!]
        \centering
    \includegraphics[width=0.9\textwidth]{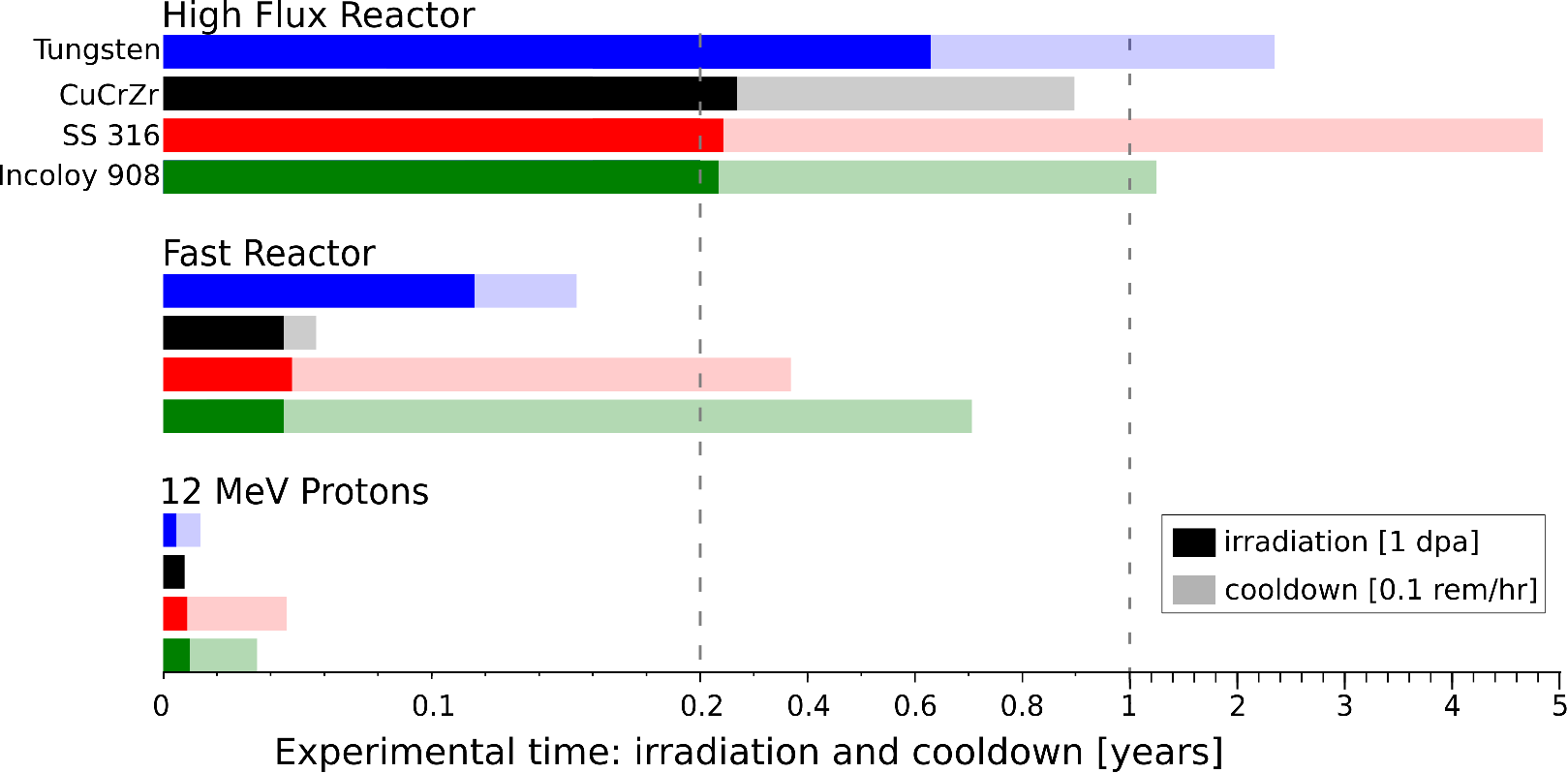}
        \caption{Damage and activation analysis demonstrates that intermediate-energy protons enable a reduction in experimental time of months to years compared to reactor irradiation}
        \label{fig:IrrTime}
\end{figure*}  

Materials were chosen to be representative of a wide span of engineering-relevant materials for future nuclear applications.  Two structural materials - a stainless steel and a nickel alloy - and two functional, high-heat-flux materials - a high-purity tungsten and a precipitate-hardened copper alloy - were selected as a subset of relevant materials.  Material compositions were taken from published work on fusion reactor relevant materials:  compositions for ITER-grade (IG) stainless steel 316, ITER-grade CuCrZr, and ITER-grade tungsten were each taken from ITER materials handbooks~\cite{Davis1996}, using the maximum allowable impurity levels. The composition for Incoloy 908 was taken from a data handbook published by the Plasma Science and Fusion Center at Massachusetts Institute of Technology~\cite{Toma1994}, using maximum allowable impurity limits.  Each of these materials has tightly controlled impurity limits in order to be minimally activating.  Material compositions are summarized in table~\ref{Tab:comp}.

The FISPACT simulations yielded data representing the radioactivity of irradiated samples while those samples decay after irradiation.  Specifically, this analysis extracted the dose to a human 30 cm away from 1~g of irradiated material, and scaled those results by representative volumes and densities.  A one-to-one comparison of neutrons and protons used results scaled by the material density and the volume of an SSJ3 tensile specimen (33~mm$^3$), which a common miniature specimen used in neutron irradiation experiments~\cite{Gussev2014}. A second set of proton results were scaled by the size of the irradiated region in a proton irradiated tensile specimen (1.5~mm$^3$), as described in section~\ref{Sec:Validate}, representing a realistic reduction in volume that can be achieved with proton irradiation.

\subsection{Radioactivity case-study for proton and neutron comparison}

The radioactivity curves in figure~\ref{fig:Activation} contain an equal-volume comparison between the two reactor scenarios (HFR and FBR) and the proton scenario, which demonstrate that radioactivity is very material-specific.  In the incoloy and copper alloy cases, protons performed similarly to high flux reactors on practical experimental time scales (days to months) In the steel case, protons generated much more radioactivity than the high flux reactors, while in the tungsten case the reverse was true.  Similarly, fast reactors allow a large decrease in radioactivity in the copper and tungsten cases, but a much more modest decrease in the steel and Incoloy cases. 

Figure~\ref{fig:Activation} also displays the benefit of reducing the proton sample volume. As seen by the darker lines, a reduced-volume proton case outperforms the full volume HFR radioactivity in every material and across the full range of times scales.  This reduced volume case even outperforms the FBR in every material but the copper alloy, where both cases represent relatively low levels of activation.

 In order to quantify the potential for time savings in a proton irradiation experiment, irradiation times and cool-down times was extracted from the FISPACT and TRIM simulations, and plotted in figure~\ref{fig:IrrTime}.  In this figure, the irradiation time to a fixed dose (1~DPA) is plotted along with the cool-down time required to bring a sample below the Nuclear Regulatory Commission definition of a high radiation area (0.1 Rem/h at 30 cm away from the source)~\cite{NRC_doserate}.  This represents the amount of time needed to irradiate and wait for sample cool-down without regulations requiring that the irradiated materials remain in strictly controlled areas during post-irradiation examination.  Additionally, this is the level of radioactivity in which a worker would reach their yearly maximum exposure (5 rem~\cite{NRC_dose}) within 50 h of being 30 cm away from a sample.  Therefore, this doserate is a representative value for high radioactivity that presents difficulties during experimentation  
 
 Figure~\ref{fig:IrrTime} demonstrates a distinct advantage in experimental time required for proton irradiation compared to reactor irradiation.  High flux reactors require extend irradiation times, requiring months to reach single-DPA doses, and creating high levels of radioactivity in samples that require years to decay away. FBRs allow a significant acceleration relative to HFRs, but still require weeks of irradiation to reach 1 DPA and months of cooldown in the steel and nickel alloy cases. Protons further reduce the irradiation time to days, and reduced sample volumes shorten the cooldown times across all four materials. 
 
 \subsection{Discussion}
The above activation analysis compares radioactivity \\across a representative set of materials and irradiation scenarios. There is not an inherent advantage in reduced activity present from the use of intermediate-energy protons. While there can be a distinct advantage in some materials (e.g. tungsten), there can also be a distinct disadvantage in others (e.g. stainless steel), when compared across equal volumes.  However, this equal volume comparison presupposes a sample size that has been optimized for use in reactors, not in proton experiments.

When activation is compared using a test geometry that has been customized for a set of proton experiments, there is a nearly universal decrease in activity across all materials. This contrast is especially stark when compared to high flux reactors, a prevalent tool for use in materials irradiation testing that can be accessed in North America~\cite{HFIR}, Europe~\cite{HFR,HFR2}, and Asia~\cite{HFANARO}. Fast reactors represent a best case-scenario for reactor irradiation and are a faster, lower-activation alternative to high flux reactors. However fast test reactors are much less common~\cite{Bor60,CEFR}, with none currently operating outside of Asia and Russia.  Even compared to fast reactors, protons offer a considerable acceleration in radiation damage and - with custom test geometries - a further reduction in activity and required cool-down times. Therefore, proton irradiation with custom test geometries has advantages in reducing the cost, time, and risk associated with radioactivity generated from materials irradiation. 

\section{Intermediate-energy proton irradiation demonstration}
\label{Sec:Validate}

Sections~\ref{Sec:PKA} -~\ref{Sec:Activation} establish principles by which intermediate-energy proton irradiation can be used as a rapid and flexible tool for bulk, radiation-damage testing.  This section provides an initial demonstration of bulk, intermediate-energy proton irradiation and the extraction of irradiated material mechanical properties through tensile testing of a metal relevant for nuclear energy systems: Alloy 718. 

\subsection{Experimental method for proton irradiation}

The experiment comprises the irradiation of Alloy 718, a high strength nickel, with 12 MeV protons followed by tensile testing. Irradiation doses and temperatures were controlled to replicate those used in a similar set of neutron irradiations performed in a high flux reactor~\cite{Byun2003}. Solution annealed Alloy 718  was chosen for its sensitivity to low levels of radiation damage and the availability of irradiated tensile test data. Solution annealed alloy 718 shows substantial hardening at doses as low as 6$\times$10$^{-4}$~DPA with yield strength increases of 100s of MPa~\cite{Byun2003}. The  compositions of the Alloy 718 varied slightly this experiment and the referenced neutron irradiations, as shown in Table~\ref{Tab:inc}.

\begin{table*}[]
    \centering
    \begin{tabular}{l|l|l|l|l|l|l|l|l|l|l|l|l|l|l}
        &Ni     & Fe    & Cr    & Nb    & Mo    & Ti    &  Al   & Co    & Mn    
        & Si    &Cu     &  C    &S      &ref\\
\hline
Proton       & 52.5  &18.5   &19     &5.13*  &3.05   &0.9    &0.5    &       &0.18
        &0.18   &0.15   &0.04   &0.008  &~\cite{Goodfellow}\\
Neutron       &  bal  &18.3   &18.13   &5.07   &3.0    & 1.1   &0.54   &0.4    & 0.21 
        &0.13   &       & 0.05  &       &~\cite{Byun2003}\\
\multicolumn{12}{l}{}\\
\multicolumn{12}{l}{*Nb + Ta}
    \end{tabular}
    \caption{Material compositions for proton and neutron irradiation experiments}
    \label{Tab:inc}
\end{table*}

Proton irradiation was performed with 12 MeV protons produced an Ionetix ION-12SC cyclotron \cite{Vincent2016}. Samples were actively water cooled during irradiation to control the temperature; sample temperature was directly monitored by thermocouples mounted to the sample surface. Irradiation temperature was kept between 80\degree C and 100\degree C and was controlled by varying the intensity of the proton beam onto the sample. The sample was irradiated to a dose of 3$\times$10$^{-4}$~DPA uniformly throughout the bulk of the sample's tensile test specimen gauge region. Details of the experimental equipment - including the cyclotron, irradiation sample, and instrumentation of the sample during irradiation - as well as the processes and irradiation conditions can be found in a previous publication~\cite{Jepeal}. 

This experiment is compared to neutron irradiations performed in the High Flux Isotope Reactor (HFIR) at Oak Ridge National Laboratory.  The original publication indicates that the irradiation temperature for these neutron irradiations was expected to be between 60\degree C and 100 \degree C~\cite{Byun2003}.

\subsection{Bulk property measurement: tensile test}
\textit{}
A post-irradiation tensile test of the Alloy 718 specimen was performed to extract the the change in strength and ductility resulting from irradiation-induced changes to the material microstructure. These tensile tests allow the extraction of a full stress-strain curve, which includes information such as the yield strength, ultimate tensile strength, ductility, and toughness. Tests were performed at room temperature, using a constant displacement rate, and with a  digital-imaging-based strain measurement system. Details on the tensile testing system and technique can be found in~\cite{Jepeal}.

\begin{figure}
    \centering
    \includegraphics[width=0.48\textwidth]{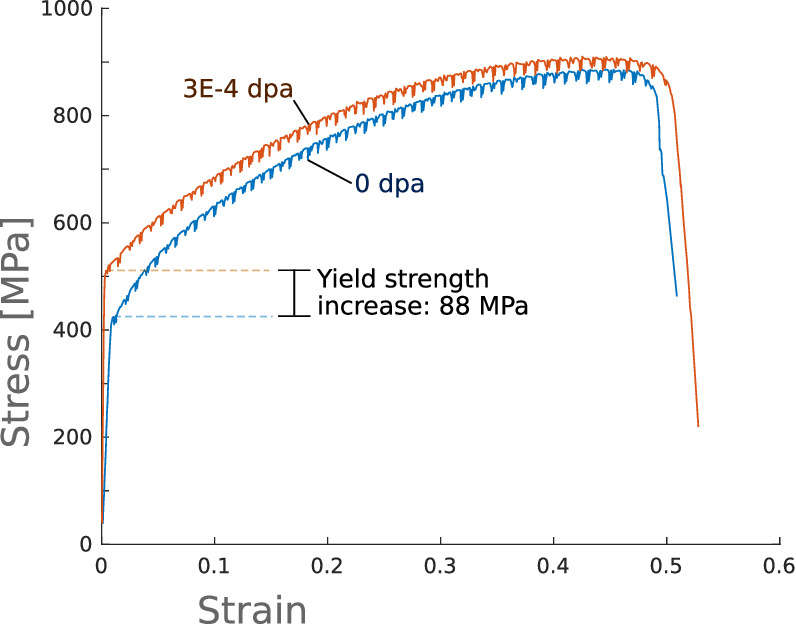}
    \caption{Low dose irradiation of a nickel-alloy highlights the sensitivity of this facility to radiation-induced property changes }
    \label{fig:StressStrain}
\end{figure}{}

The results of the tensile test are shown in Figure~\ref{fig:StressStrain}, which shows stress-strain curves for two Alloy 718 samples, one pristine and one irradiated to 3$\times$10$^{-4}$~DPA. The irradiated samples shows expected changes in the stress strain curve indicative of radiation damage induced alteration of the material microstructure. In particular, the yield strength of the material increased by 20 percent, from 424 MPa in the pristine case to 511 MPa in the irradiated case, which quantitatively matches results in previous irradiation studies of Alloy 718 at these dose levels. Importantly, the uncertainty in the yield strength measurement with the specimen geometry and tensile tester used for this experiment is less than 10~MPa \cite{Jepeal}, well below the measured yield strength change at such low radiation dose. This provides confidence that the resulting yield strength change is a direct result of irradiation and not due to the equipment or uncertainty. This experiment also demonstrates the high sensitivity to material property changes of the technique even at relatively low radiation doses in materials like Alloy 718. Additionally, there is no significant change in ultimate tensile strength and ductility are observed in this sample, consistent with previous studies of irradiated Alloy 718 at such low doses.

An important final result from this experiment is how closely the hardening of this sample induced by \textit{12 MeV protons} qualitatively and quantitatively matches the hardening induced by \textit{reactor spectrum neutrons}~\cite{Byun2003}. Preliminary irradiations with 12~MeV protons to higher doses indicate similarly close changes in other mechanical properties, such as uniform ductility and ultimate tensile strength. This replication of neutron irradiation-induced material property changes with 12~MeV protons suggests that the microstructural changes to the material in both neutron and proton cases are highly similar. More detailed irradiations are now being carried out on a larger selection of materials and to a wider range of total doses. Macroscopic testing such as tensile tests for engineering material properties are being directly compared with past reactor irradiation experiments, and this will be presented in an upcoming publication. If close comparison are found, this would further confirm the capability of IEPI to produce material changes with high-fidelity to reactor environments, as discussed above in Section~\ref{Sec:PKA}. The use of IEPI would then provide a rapid, high fidelity tool to achieve understanding, down selection, and qualification of materials for advanced nuclear energy systems. 

\section{Conclusion: impact on nuclear design}
\label{sec:Impact}

In this article, intermediate-energy protons have been identified as a uniquely flexible tool for studying radiation damage for nuclear power technology.  Unlike most other ion-based techniques, 10 -- 30 MeV protons are capable of damaging samples with thicknesses of hundreds of \SI{}{\micro\meter}, enabling bulk property measurement. The recoil energies produced by these protons are well matched to those of fission and fusion reactor neutrons, ensuring similar radiation damage cascades and primary damage production. The production of helium and hydrogen through transmutation is controllable based on the incident proton energy, and can be matched to any of the wide range of neutron environments.  This combination of bulk irradiation, representative recoil energies, and controllable transmutations cannot be found in any other irradiation technique.

Like other ion irradiation techniques, IEPI is also a fast, low-cost tool for radiation damage testing. With exposure-rates that are many times higher than test reactor conditions, protons allow time savings of months or years compared to reactor irradiation.  The ability to reduce sample activation leads to a similar  time saving when waiting for irradiated materials to “cool-down,” as well as reducing the need for costly radioactive testing equipment.  With new accelerator technology, proton experiments are now being performed in a small university laboratory with minimal personnel (1-2 persons) and equipment operation and maintenance costs ($\sim$\$10k/year)~\cite{Jepeal}, in stark contrast to the tens of millions of dollars needed to staff, fuel, and operate research reactors~\cite{DOE2020}.  

The cost (millions of dollars) and length (several years) of reactor irradiation and testing campaigns pose a major challenge for first-of-a-kind nuclear systems,  such as plasma facing components for fusion power plants or internal structures for a molten salt reactor. Designers of these systems are, in practice, allowed only a single experimental step for validating the radiation resistance of their materials, leaving no room for implementing the insights from these irradiation campaigns through iterative design improvement. With a faster, lower-cost tool, developers can iterate design with multiple stages of radiation damage experimentation. This allows, for instance, rapid screening of many (tens) of material variations (compositions, manufacturing methods, heat treatments, etc.), followed by in depth characterization of the most promising candidates from screening. As the potential pool of materials are narrowed, designers can continue using the results of early experiments to inform better allocation of the later experiments. Key phenomena like the onset of void swelling or a radiation-induced transition from ductility to brittleness are currently very coarsely approximated by single-pass reactor irradiation; with a fast, iterative technique, the critical doses for the onset of these phenomena could be resolved as finely as needed. Through iterative design-experimentation cycles like this, IEPI could allowing finite resources to be much more efficiently allocated for design.  

With high-fidelity, faster evaluation of radiation damage effects, and the unlocking of iterative design processes, IEPI offers the tools to answer essential questions for the design of new nuclear systems. For instance, the design of a tungsten plasma facing component requires assessing the impact of irradiation embrittlement on the component lifetime; a designer can now generate predictive data across a range of potential system designs and material choices, where they would previously depended on the sparse test reactor data that already exists. With this information, comes the ability to evaluate key trade-offs between system capital costs, maintenance needs, and lifetime, which are essential to the economics of these systems. Therefore, intermediate energy proton irradiation could profoundly improve the design and commercial viability of advanced nuclear energy technologies.

\section*{Acknowledgements}
The authors are grateful to the support and suggestions made by Michael Short, Cody Dennett, and Leigh Ann Kesler, as well as the guidance and management of  Alberto Pontarollo, Maria Elena Gennaro, and their colleagues at Eni.

This work was funded by Eni S.p.A. through the MIT Energy Initiative and the Laboratory for Innovation in Fusion Technology (LIFT) at the MIT Plasma Science and Fusion Center.

\section*{Data Availability}
The raw data and processed data required to reproduce these findings are available to download from \url{http://dx.doi.org/10.17632/ndj55thg39.1}

\bibliography{main}

\end{document}